\documentclass[a4paper,fleqn]{style}
\usepackage{todonotes}
\usepackage{xcolor}

\usepackage[numbers]{natbib}

\def\tsc#1{\csdef{#1}{\textsc{\lowercase{#1}}\xspace}}
\tsc{WGM}
\tsc{QE}

\begin{document}
\let\WriteBookmarks\relax
\def\floatpagepagefraction{1}
\def\textpagefraction{.001}

\shorttitle{}    

\shortauthors{}  

\title [mode = title]{Spin‑wave hybridization in bismuth iron garnet Mie spheres induced by the inverse Faraday effect}  

\tnotemark[1] 

\tnotetext[1]{} 

%

\author[1]{Fedor Shuklin}

\cormark[1]


\ead{fedor.shuklin.95@mail.ru}


\credit{
Conceptualization of this study, Formal analysis, Methodology, Software, Writing – original draft, Visualization, Data curation
}

\affiliation[1]{organization={Center for Photonics and 2D Materials, Moscow Institute of Physics and Technology},
            addressline={Institutskiy per., 9}, 
            city={Dolgoprudny},
            postcode={141700}, 
            state={Moscow Region},
            country={Russia}}

\author[2]{Khristina Albitskaya}

\affiliation[2]{organization={Moscow Center for Advanced Studies},
            addressline={Kulakova str. 20}, 
            city={Moscow},
            postcode={123592}, 
            country={Russia}}

\credit{Visualization, Validation, Writing – review and editing, Conceptualization}

\author[1,3]{Alexander Chernov}

\affiliation[3]{organization={Russian Quantum Center},
            addressline={Bolshoi Boulevard, Building 30,  Skolkovo Innovation Center}, 
            city={Moscow},
            postcode={121205}, 
            country={Russia}}

\credit{Writing – review and editing, Conceptualization, Supervision, Project administration}

\author[4]{Mihail Petrov}
\credit{Writing – review and editing, Conceptualization, Validation, Supervision, Project administration}

\affiliation[4]{organization={Department of Physics and Engineering, ITMO University},
            addressline={Lomonosova St. 9}, 
            city={St. Petersburg},
            postcode={191002}, 
            country={Russia}}


\begin{abstract}
We show that the inverse Faraday effect can be used to engineer dipole--exchange spin-wave spectra in ferrimagnetic bismuth iron garnet (BIG) Mie spheres. Internal optical Mie resonances generate spatially structured effective magnetic fields whose symmetry is inherited from the optical near field and which act as controllable perturbations of the magnon Hamiltonian. For circularly polarized light incident collinearly with the equilibrium magnetization, the optical perturbation preserves axial symmetry while breaking mirror parity, thereby enabling hybridization of magnon modes with opposite parity within the same $\widehat{J}_z$ sector. Using coupled-mode theory, we derive the corresponding avoided-crossing spectrum and analytical expressions for the induced level splittings, which scale linearly with pump intensity. Numerical calculations for BIG spheres confirm the predicted hybridization and show that the splitting is maximized near optical Mie resonances, where field enhancement and magneto-optical response are strongest. We further discuss the roles of damping, linewidth, and heating, and show that the predicted MHz--hundreds-of-MHz splittings should be observable under realistic conditions. These results identify BIG Mie resonators as a promising platform for symmetry-selective optical control of spin-wave spectra.
\end{abstract}




\begin{keywords}
optomagnonics \sep inverse Faraday effect \sep spin waves \sep Mie resonances \sep bismuth iron garnet \sep mode hybridization \sep coupled-mode theory
\end{keywords}
\maketitle
\section{Introduction}\label{sec:introduction}

Optomagnonics, the study of the interaction between optical photons and spin-wave excitations in magnetic media, has emerged as a route toward efficient coupling between GHz magnonic signals and optical carriers on integrated platforms \cite{kusminskiy2019cavityoptomagnonics, Wang_Gao_Jiao_Wang_2022, nano5020577}. A central mechanism in this context is the inverse Faraday effect (IFE), whereby circularly polarised light generates an effective magnetic field capable of driving magnons without significant heating \cite{van1965optically}. Owing to this capability, the IFE has become a versatile tool for accessing and controlling magnetization dynamics with light \cite{Wang_Gao_Jiao_Wang_2022}. Its implementation has progressed from extended magneto-optical structures to increasingly confined photonic platforms, including resonant waveguides \cite{zhu2022inverse,pantazopoulos2020planar, Valagiannopoulos2007PIER}, dielectric nanoresonators \cite{PhysRevApplied.22.064087}, and, more recently, subwavelength Mie scatterers and metasurfaces \cite{fan2019magneto, 10.1063/1.5066307, 10.1063/5.0257020, Katsantonis2023SciRep}. This evolution reflects the broader effort to enhance light--magnon coupling through stronger optical confinement and improved spatial overlap with spin-wave modes.

\begin{figure}
  \centering
    \includegraphics[width=1\columnwidth]{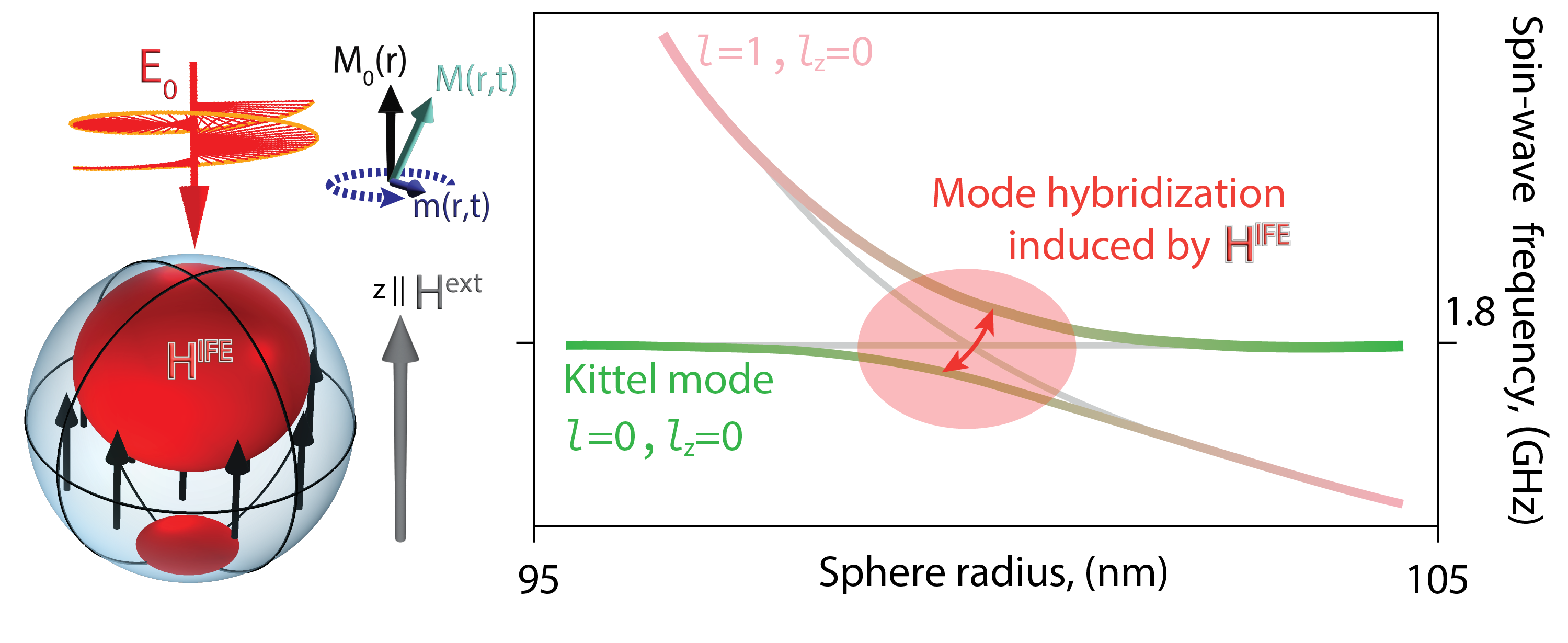}
    \caption{Schematic evolution of spin‑wave modes in a ferromagnetic sphere driven by circularly polarized light $E_{\text{0}}$ as the system radius grows. The inverse Faraday effect produces hybridization and avoided crossings between the $l=0$ Kittel mode and higher‑order odd mode with $l=1$.}
    \label{fig:Figure_1}
\end{figure}

Confined resonators are especially attractive because they amplify optomagnonic interactions \cite{PhysRevB.101.054412}. In whispering-gallery and Fabry--P\'erot systems, discrete optical modes can couple efficiently to magnons via magneto-optical effects, enabling enhanced Brillouin scattering and coherent photon--magnon conversion \cite{PhysRevLett.117.123605, Kozhaev2018,PhysRevB.97.214423}. The emerging regime of Mie optomagnonics pushes this miniaturization further: submicron scatterers support multipole resonances with structured near fields that can endow the IFE with nontrivial spatial texture and angular momentum \cite{PhysRevB.101.054412,PhysRevB.104.214429,PhysRevApplied.22.064087}. These resonances not only enhance the internal field, but also provide additional control through wavelength-dependent nodal structure and mode symmetry \cite{PhysRevApplied.22.064087}.

So far, optical fields in optomagnonic structures have been used primarily either to excite magnons or to probe them through Brillouin scattering~\cite{Chernov:17,PhysRevB.97.214423}. 
In particular, spatially nonuniform inverse-Faraday fields in magnetophotonic and optomagnonic microcavities have enabled selective excitation of standing spin waves of different orders \cite{Ozerov2022JMMM,Krichevsky2024Nanophotonics}, while integrated optomagnonic waveguides have demonstrated coherent optical launching of spin waves at low optical powers~\cite{Zhu2022PRApplied}. 
These results show that structuring the optical field provides a powerful handle on the efficiency and selectivity of magnon excitation. 
A more demanding and less explored possibility is to use light not only as a source or probe of spin waves, but as a controllable perturbation of the magnon Hamiltonian itself, thereby modifying mode frequencies, hybridization, spatial structure, and dissipation \cite{10.1063/1.5100826,Zhang2026}.

Realizing such control requires accounting for the nontrivial spin-wave dynamics of nanoscale resonators, where exchange and dipolar interactions compete and the relevant modes often lie in the intermediate dipole--exchange regime.
In this respect, the recently developed symmetry-based coupled-mode framework for confined magnetic particles is particularly useful, because it identifies the conserved quantities---namely, the $z$-projection of the total angular momentum, $\widehat{J}_z$, and the $xOy$-plane reflection parity, $\widehat{P}_z$---that organize the dipole--exchange spectrum and therefore determine which perturbation-induced couplings are allowed or forbidden~\cite{shuklin2026dipoleexchangespinwavesmode}.

At the same time, magnetic nanospheres and related curved particles can exhibit rich spin-wave spectra even without optical driving.
Static geometry, curvature, vortex or Bloch-point textures, surface asymmetry, and multilayered or core--shell morphology can shift resonance frequencies, mix modes, and activate otherwise weak or forbidden resonances~\cite{Kim2015SciRep,McKeever2019PRB, McKeever2022ACSAMI,Saavedra2024ACSAMI}. 
These mechanisms are important for understanding magnetic nanoparticles, but they are intrinsically static: the coupling is fixed by the fabricated geometry, equilibrium magnetization texture, or material profile. In contrast, the mechanism considered below is induced dynamically by the optical pump.
The Mie-resonant inverse-Faraday field acts as a spatially structured Zeeman-like perturbation whose symmetry can be controlled by the wavelength, intensity, polarization, and incidence geometry of light.
This provides a route toward real-time, reconfigurable control of dipole--exchange spectra in homogeneous, uniformly magnetized ferrimagnetic particles.

Building on this symmetry framework, we use the effective IFE fields of Krichevsky \textit{et al.}~\cite{PhysRevApplied.22.064087} to demonstrate symmetry engineering of spin-wave spectra in ferrimagnetic bismuth iron garnet (BIG) Mie spheres, as sketched in Fig.~\ref{fig:Figure_1}.
BIG is particularly advantageous in this context because, compared with YIG, it provides a substantially stronger magneto-optical response while still retaining sufficiently low optical loss near the chosen pump wavelength, which is fixed at $582\,\mathrm{nm}$ throughout the paper \cite{Wang_Gao_Jiao_Wang_2022,Yao_2015}. 
We show that for circularly polarized light incident collinearly with the equilibrium magnetization, the optical perturbation preserves axial symmetry, so that $\widehat{J}_z$ remains conserved and the exchange label $l_z$ remains a useful approximate identifier in this sector, while mirror parity is broken. 
This enables hybridization of magnon modes with opposite parity within the same $l_z=0$ sector---specifically, the even Kittel mode with orbital angular momentum $l=0$ and the first odd mode with $l=1$---as illustrated in Fig.~\ref{fig:Figure_1}. 
Using coupled-mode theory, we derive the resulting avoided crossing and the corresponding level splitting, which scales linearly with the pump intensity, $\propto |E_0|^2$. Finally, we discuss the experimental feasibility of observing this effect, including the roles of finite linewidth and optical heating.

\section{Spin-wave dynamics and inverse-Faraday coupling}

\begin{figure*}
\label{fig:Figure_2}
  \centering
    \includegraphics[width=1\textwidth]{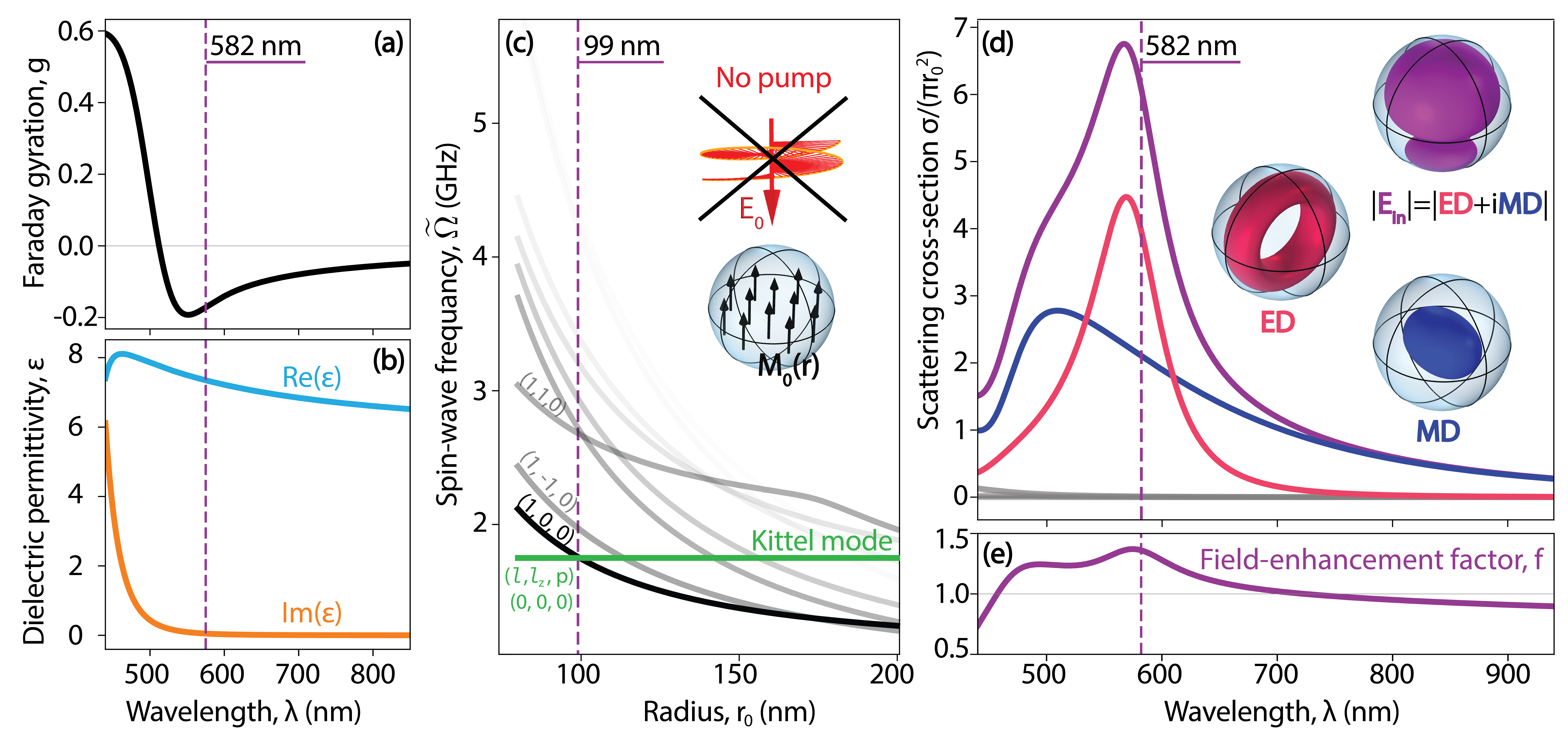}
    \caption{Optical and magnetic properties of a BIG sphere: dispersion of (a) Faraday gyration constant and (b) real (light-blue line) and imaginary (orange line) parts of dielectric permittivity; dashed line in (a) and (b) shows chosen pump wavelength of $582 \ \mathrm{nm}$; (c) spin-wave spectra of unperturbed (no pump) sphere with green line showing Kittel $l=0, l_z=0, p=0$ mode, and solid black line depicting odd $l=1,l_z=0,p=0$ mode; dashed line in (c) marks crossing radius of $99 \ \mathrm{nm}$; inset in (c) shows equilibrium uniform magnetization field (black arrows); figure (d) presents normalized total scattering cross-section (purple line) and Electric dipole (ED, pink line) and Magnetic dipole (MD, blue light) components; insets in (d) show iso-surfaces of electric field amplitude of ED and MD; (e) shows electric field enhancement factor as defined in Eq.\eqref{eq:enhancement-factor}.}
\end{figure*}

We consider a ferrimagnetic nanosphere of bismuth iron garnet (BIG), magnetized to saturation by an external static field $\mathbf{H}^{\mathrm{ext}}=5\cdot10^{5}\si{A / m}$ and $\mathbf{H}^{\mathrm{ext}}\parallel \hat{\mathbf z}$, so that the equilibrium magnetization is uniform, $\mathbf{M}_0=M_s\hat{\mathbf z}$, where $M_s= 1.3\cdot10^5\,\si{A/ m}$ \cite{Yao_2015} is the saturation magnetization. 
A sketch of the system is shown in Fig.~\ref{fig:Figure_1}, where the black arrows indicate the uniform magnetization and the direction of the external magnetic field.
Spin waves in this system are governed by the Landau--Lifshitz--Gilbert equation
\begin{equation}
\partial_t \mathbf{M}
=
-\gamma\,\mathbf{M}\times \mathbf{H}^{\mathrm{eff}}
+
\frac{\alpha}{M_s}\,
\mathbf{M}\times \partial_t \mathbf{M},
\label{eq:LLG_full}
\end{equation}
where $\gamma$ is the gyromagnetic ratio and $\alpha$ is the Gilbert damping constant. In the linear regime we write
\begin{equation}
\mathbf{M}=\mathbf{M}_0+\mathbf{m},
\qquad
|\mathbf{m}|\ll M_s,
\end{equation}
which yields
\begin{align}
\partial_t \mathbf{m}
&=
\gamma\bigl(
\mathbf{H}_0^{\mathrm{eff}}\times \mathbf{m}
-
\mathbf{M}_0\times \mathbf{h}[\mathbf{m}]
\bigr)- 
\nonumber\\
&\quad
\frac{\alpha\gamma}{M_s}\,
\mathbf{M}_0\times
\bigl(
\mathbf{H}_0^{\mathrm{eff}}\times \mathbf{m}
-
\mathbf{M}_0\times \mathbf{h}[\mathbf{m}]
\bigr).
\label{eq:LLG_lin}
\end{align}
Here $\mathbf{H}_0^{\mathrm{eff}}\parallel \hat{\mathbf z}$ is the effective stationary field, containing the external field and the static demagnetizing contribution $\mathbf{H}_0^{\rm{dm}}=-1/3 M_s\mathbf{\hat{z}}$, while $\mathbf{h}[\mathbf{m}]$ is the dynamical field induced by the magnetization perturbation. In an undriven setup dynamical field contains only demagnetizing field $\mathbf{h}^{\rm{dm}}[\mathbf{m}]$ and an effective exchange field
\begin{equation*}
    \mathbf{h}^{\rm{exch}} = \frac{A}{M_s}\nabla^2\mathbf{m},
\end{equation*}
with $A=4.28\cdot10^{-11} \si{A\cdot m}$ \cite{10.1063/1.1287227} being an exchange stiffness, and $\nabla^2$ is the vector Laplacian.
In what follows, we first consider the lossless problem, $\alpha=0$, and return to damping only later when discussing linewidths and splittings.

For a uniformly magnetized sphere, it is convenient to pass directly to the operator formulation of Ref.~\cite{shuklin2026dipoleexchangespinwavesmode}, in which the lossless linearized dynamics takes the form

\begin{equation}
-i\partial_t \mathbf m(\mathbf r,t)
=
\bigl(
\mathcal H_{\rm exch}
+
\mathcal H_{\rm dip}
\bigr)\mathbf m(\mathbf r,t).
\label{eq:H_total}
\end{equation}
The exchange contribution is the local operator
\begin{equation}
\mathcal{H}_{\rm{exch}} = \gamma \Bigl( \ell_{\rm exch}^2 M_s \nabla^2 - H_0^{\rm eff} \Bigr) \hat{S}_z,
\label{eq:H_exch}
\end{equation}
with
\begin{equation*}
\ell_{\rm{exch}}=\sqrt{\frac{A}{M_s}},
\qquad
H_0^{\rm{eff}}
=
H^{\rm{ext}}-\frac{1}{3}M_s.
\label{eq:lexch_Heff}
\end{equation*}
The dipolar contribution is the nonlocal operator
\begin{equation}
\mathcal H_{\rm dip}\,\mathbf m
=
\gamma M_s \hat S_z
\int_V d^3r'\,
\hat K(\mathbf r-\mathbf r')\,\mathbf m(\mathbf r'),
\label{eq:hdip_def}
\end{equation}
defined through the magnetostatic kernel
\begin{equation}
\hat K(\mathbf r-\mathbf r')
=
-\nabla\nabla
\frac{1}{4\pi |\mathbf r-\mathbf r'|}.
\label{eq:K_def}
\end{equation}


In this framework, the spin‑wave spectrum is organized by the symmetries of the Hamiltonian, which were fully analyzed in Ref.~\cite{shuklin2026dipoleexchangespinwavesmode}. In the exchange limit, the spherical modes can be expressed rigorously and labeled by $(l,l_z,p)$, with $l$ the orbital angular momentum, $l_z$ its projection on the $z$ axis, and $p$ the radial quantum number. In the full dipole--exchange regime, however, the conserved quantities reduce to the total angular-momentum projection $\widehat{J}_z$ and the mirror parity $\widehat{P}_z$, so the exchange labels remain only approximate, though still useful prior to strong hybridization. At the same time, the analytic exchange solutions furnish the natural basis for the coupled-mode theory, allowing one to describe dipolar-induced hybridization and to identify from symmetry which intermode couplings are allowed or forbidden.

The optical perturbation to the spin-wave system enters via the inverse Faraday effect (IFE), governed by the optical spin density \cite{zhu2022inverse}
\begin{equation*}
\mathbf s=-i\,\mathbf E_{\text{in}}^\ast\times \mathbf E_{\text{in}}.
\label{eq:s_opt}
\end{equation*}
The electric field $\mathbf E_{\text{in}}$ generating this optical spin is the internal Mie field excited inside the sphere by a circularly polarized plane wave of helicity $\sigma=\pm1$ and amplitude $E_0$. In the simplest geometry, in which the pump propagates collinearly with $\mathbf M_0\parallel \hat{\mathbf z}$, this field admits the multipole expansion~\cite{PhysRevApplied.22.064087}
\begin{equation}
\mathbf{E_{\text{in}}}(\mathbf{r})
=
E_0
\sum_{L=1}^{\infty}
q_L
\left[
c_L\,\mathbf{M}_{m L}(\kappa\mathbf r)
+
i d_L\,\mathbf{N}_{m L}(\kappa\mathbf r)
\right],
\label{eq:Mie_internal}
\end{equation}
where $q_L=i^L\sqrt{4\pi(2L+1)}$, $\kappa$ is the radial wavenumber inside the sphere, $c_L$ and $d_L$ are the internal Mie coefficients, and $\mathbf N_{m L}$ and $\mathbf M_{m L}$ are electric and magnetic multipoles respectively, with azimuthal index$m=\sigma$ defined by the pump helicity $\sigma$.
The corresponding effective IFE magnetic field is
\begin{equation}
\mathbf H^{\mathrm{IFE}}
=
i\,
\frac{\varepsilon'(\lambda)\varepsilon_0}{\mu_0}\,
\frac{g(\lambda)}{M_s}\,
\mathbf E_{\text{in}}^\ast\times \mathbf E_{\text{in}}.
\label{eq:H_IFE}
\end{equation}
Here $\varepsilon(\lambda)=\varepsilon'(\lambda)+i\varepsilon''(\lambda)$ is the complex dielectric function of BIG, which determines the internal Mie field distribution, while $g(\lambda)$ is the dispersive magneto-optical gyration parameter with the dispersion obtained from Yao \textit{et al.} \cite{Yao_2015} shown in Fig.~\ref{fig:Figure_2}(a), that sets the strength of the IFE coupling. 
BIG is particularly favorable in this context because it combines strong and strongly wavelength-dependent Faraday response with comparatively low optical absorption and small dispersion above $\sim 500\,\mathrm{nm}$, as can be seen in Fig.~\ref{fig:Figure_2}(b) allowing resonant optical enhancement without excessive heating~\cite{Yao_2015}.

The IFE enters the spin-wave problem as an additional contribution to the effective magnetic field,
\begin{equation*}
\mathbf H^{\mathrm{eff}}_0
\;\to\;
\mathbf H^{\mathrm{eff}}_0+\mathbf H^{\mathrm{IFE}}(\mathbf r),
\label{eq:Heff_with_IFE}
\end{equation*}
and therefore acts as an optically induced perturbation of the magnon Hamiltonian. Since the spin-wave dynamics is transverse, $\mathbf m\perp \mathbf M_0$, only the component of $\mathbf H^{\mathrm{IFE}}$ parallel to the equilibrium magnetization contributes to the linearized torque to leading order. In the present geometry, with $\mathbf M_0\parallel \hat{\mathbf z}$, this means that only the $z$-component of the IFE field is relevant, $H^{\mathrm{IFE}}_z(\mathbf r)
=
\hat{\mathbf z}\cdot \mathbf H^{\mathrm{IFE}}(\mathbf r)$,
while the transverse components do not enter the linear problem at this order.

Accordingly, within the operator formulation of Ref.~\cite{shuklin2026dipoleexchangespinwavesmode}, the IFE gives rise to an additional local term in the effective Hamiltonian,
\begin{equation}
-i\partial_t \mathbf m(\mathbf r,t)
=
\bigl(
\mathcal H_{\rm exch}
+
\mathcal H_{\rm dip}
+
\mathcal H_{\rm IFE}
\bigr)\mathbf m(\mathbf r,t),
\label{eq:H_total_IFE}
\end{equation}
with
\begin{equation}
\mathcal H_{\rm IFE}
=
-\gamma\,H^{\mathrm{IFE}}_z(\mathbf r)\,\hat S_z .
\label{eq:H_IFE_op}
\end{equation}
Thus, the inverse Faraday effect enters as a spatially structured, optically controlled Zeeman-like term whose symmetry is entirely determined by the internal Mie field profile. In particular, it is this term that can preserve or break the conserved quantities $\widehat{J}_z$ and $\widehat{P}_z$, and thereby induce or forbid coupling between different spin-wave modes.

\section{IFE‑induced coupling between spin‑wave modes}
\label{sec:collinear_incidence}

We consider a BIG sphere of radius $r_0=110\,\mathrm{nm}$, whose unperturbed dipole--exchange spin-wave spectrum is shown in Fig.~\ref{fig:Figure_2}(c). Throughout this section we use the labeling introduced in Ref.~\cite{shuklin2026dipoleexchangespinwavesmode}, in which exchange modes are denoted by triplets $(l,l_z,p)$, where $l$ is the orbital angular momentum, $l_z$ its projection on the $z$ axis, and $p$ the radial index. Although this notation is exact only in the exchange-dominated regime, it remains useful for identifying dipole--exchange modes in spectral regions where dipole coupling has not yet become strong.

We focus on the Kittel mode $(l,l_z,p)=(0,0,0)$, indicated by the green line, and the first odd mode $(1,0,0)$, shown by the solid black line. 
In the unperturbed spectrum, these modes cross at approximately $r_0=99\,\mathrm{nm}$; accordingly, below we examine the vicinity of this crossing in the radius range $(95-105)\,\si{nm}$.
Figure~\ref{fig:Figure_2}(c) shows the unpumped reference crossing: the Kittel branch and the first odd branch intersect without an avoided crossing.
This follows from the symmetry classification of Ref.~[24], since the unpumped spherical Hamiltonian preserves $\widehat{J}_z$ and $\widehat{P}_z$, so opposite-parity modes do not mix.

The insets in Fig.~\ref{fig:Figure_2}(d) show the isosurfaces of $\left|\operatorname{Re}\,\mathbf{E}_{\text{in}}\right|$ for the ED and MD modes, while Fig.~\ref{fig:Figure_2}(e) displays the field-enhancement factor, defined as the volume-averaged ratio 
\begin{equation}
    f = \frac{\langle |\mathbf E_{\mathrm{in}}|^2\rangle_V}{|E_0|^2} = \frac{1}{V|E_0|^2}\int_V{d^3\mathbf{r} \ |\mathbf{E}_{\rm{in}}|^2}.
    \label{eq:enhancement-factor}
\end{equation}
The corresponding internal optical-field profile for circularly polarized excitation is shown in the inset of Fig.~\ref{fig:Figure_2}(c).

We consider the collinear-incidence geometry, in which the circularly polarized pump propagates parallel to equilibrium magnetization, $\mathbf{k}_{\rm 0}\parallel \mathbf{M}_0\parallel \hat{\mathbf z}$. In this configuration, the optical axis coincides with the symmetry axis of the sphere, so the internal field remains axially symmetric. For a pump with helicity $\sigma=+1$, the excited Mie field contains only the $m=1$ channel. 
In the spectral region of interest, the optical response is governed primarily by the simultaneous excitation and interference of the electric-dipole (ED) and magnetic-dipole (MD) Mie channels. These two channels are both generated by the circularly polarized incident wave,
but their angular structures have opposite behavior under the reflection
$z\rightarrow -z$, so the ED contribution is proportional to $Y^1_1$ and is therefore odd under this transformation, whereas the dominant MD field is composed of harmonics with orbital indices $L=0,2$ and is therefore even. Consequently, in the bilinear optical-spin density the parity-breaking part originates from the ED--MD cross-terms rather than from a single parity channel alone. 
These cross-terms preserve axial symmetry, because the pump still selects a single azimuthal channel $m=\sigma$, but they break mirror symmetry with respect to the equatorial plane. At the same time, because the internal optical field is spatially nonuniform, the induced IFE perturbation is also nonuniform.
Together, parity breaking and spatial inhomogeneity allow the optical drive to couple magnon modes belonging to the same axial-symmetry sector but having opposite mirror parity.

\begin{figure}
\label{fig:Figure_3}
  \centering
    \includegraphics[width=1\columnwidth]{Figure_3.png}
    \caption{Results of numerical and coupled-mode calculations of: (a) hybridization of the spectral lines at pump intensity $I=5.3\times10^5\,\mathrm{W/cm^2}$; green circles show the numerical data, dashed lines show the uncoupled eigenlines obtained from the diagonal CMT Hamiltonian, and solid lines show the full coupled-mode result; (b) splitting magnitude $\Delta\Omega$ as a function of wavelength for a sphere of radius $r_0\approx 99\,\mathrm{nm}$ at pump intensity $I=5.3\times10^5\,\mathrm{W/cm^2}$; (c) splitting magnitude as a function of pump intensity for the same radius and pump wavelength $\lambda=582\,\mathrm{nm}$. In panel (c), the horizontal dashed line marks the spin-wave linewidth of $14.4\,\mathrm{MHz}$ and thus the approximate threshold for resolving the split, while the vertical dashed line indicates the corresponding pump intensity.}
\end{figure}

To demonstrate this IFE-induced hybridization, we compute the spin-wave spectra numerically in COMSOL Multiphysics 6.1. The linearized Landau--Lifshitz--Gilbert equation is solved with the Micromagnetics Module \cite{comsol, 10.1063/5.0143262}, while the demagnetizing field is obtained using the Magnetic Fields, No Currents interface. 
In the micromagnetic domain, the finite-element mesh inside the sphere was constrained so that the maximum element size did not exceed the exchange length $l_{\rm exch}$, ensuring that the shortest exchange-controlled spatial variations were resolved.
The surrounding nonmagnetic domain used for the magnetostatic calculation was chosen as a sphere of radius $3r_0$.
This value was obtained by increasing the outer-domain size and verifying convergence of the relevant eigenfrequencies.
The mesh in the outer magnetostatic domain was generated automatically by the Magnetic Fields, No Currents interface.

The results of the numerical analysis are shown by hollow-circle markers in Fig.~\ref{fig:Figure_3}. Panel (b) presents the splitting $\Delta\Omega$ as a function of wavelength and identifies the maximum splitting resulting from the interplay between the resonant optical response of the particle and the spectral dispersion of the Faraday gyration constant. Panel (c) shows the dependence of the splitting on the pump intensity, $I=\tfrac{1}{2}\varepsilon_0 c|\mathbf E_0|^2$, demonstrating linear behavior. Panel (a) displays the corresponding zoomed-in avoided crossing, as schematically illustrated in Fig.~\ref{fig:Figure_1}(b).

To gain further insight into the hybridization mechanism, we employ magnonic coupled-mode theory (CMT) following Ref.~\cite{shuklin2026dipoleexchangespinwavesmode}, using the exchange eigenmodes of the sphere as the natural basis. In the exchange limit, the positive-frequency branch is
\begin{equation}
m_{-,\nu}(\mathbf r)=
C_{l l_z p}\,
j_l\!\left(\kappa_{lp}\frac{r}{r_0}\right)
Y_l^{\,l_z}(\theta,\phi),
\label{eq:exchange_mode_collinear}
\end{equation}
where $\nu\equiv(l,l_z,p)$, $j_l$ is the spherical Bessel function, $Y_l^{\,l_z}$ is the spherical harmonic, and $\kappa_{lp}$ is determined by the exchange boundary condition $j_l'(\kappa_{lp})=0$. The corresponding full vector mode is written as
\begin{equation}
\mathbf m_\nu(\mathbf r)=m_{-,\nu}(\mathbf r)\,\mathbf e_-,
\end{equation}
where $\mathbf e_-$ is the circular-polarization vector of the positive-frequency branch.

Treating the inverse-Faraday field as a perturbation, we write the linearized spin-wave dynamics as
\begin{equation}
-i\partial_t \mathbf m(\mathbf r,t)=
\left(
\mathcal H_{\rm dip\text{-}exch}
+
\mathcal H_{\rm IFE}
\right)\mathbf m(\mathbf r,t),
\end{equation}
where $\mathcal H_{\rm dip\text{-}exch}$ is the dipole--exchange Hamiltonian of the unpumped sphere, whose eigenfrequencies are denoted by $\widetilde\Omega_\nu$. Expanding the dynamical magnetization over the two relevant exchange modes introduced in Eq.~\eqref{eq:exchange_mode_collinear},
\begin{equation}
\mathbf m(\mathbf r,t)=c_{000}(t)\,\mathbf m_{000}(\mathbf r)+c_{100}(t)\,\mathbf m_{100}(\mathbf r),
\end{equation}
and projecting onto this truncated basis yields
\begin{equation}
\begin{split}
i\dot c_{000}&=\widetilde\Omega_{000}\,c_{000}
+U_{000}\,c_{000}
+U_{000}^{100}\,c_{100},\\
i\dot c_{100}&=\widetilde\Omega_{100}\,c_{100}
+U_{100}\,c_{100}
+\left(U_{000}^{100}\right)^\ast c_{000},
\end{split}
\label{eq:CMT_collinear_time}
\end{equation}
where $\widetilde\Omega_{000}$ and $\widetilde\Omega_{100}$ are the dipole-corrected eigenfrequencies of the unpumped sphere.
The two-mode truncation is justified by Fig.~\ref{fig:Figure_2}(c): within the studied radius window the nearest additional branch belongs to a different $j_z$ sector, while the collinear IFE perturbation preserves axial symmetry and can only require a multimode CMT if another optically allowed mode becomes nearly resonant within the scale of the IFE coupling or linewidth.
In the absence of optical driving, hybridization between the Kittel mode $(0,0,0)$ and the mode $(1,0,0)$ is symmetry-forbidden, so the dipolar interaction contributes only through these diagonal frequency shifts and does not generate intermode coupling.
In the two-mode CMT basis, the unpumped dipole--exchange Hamiltonian is diagonal for this pair of modes.
The dashed lines in Fig.~\ref{fig:Figure_3}(a) show the corresponding crossing when the IFE-induced off-diagonal term $U^{100}_{000}$ is omitted.
Consistently, Fig.~\ref{fig:Figure_3}(c) shows $\Delta\Omega\rightarrow 0$ as $I\propto |E_0|^2\rightarrow 0$.

The inverse Faraday effect produces additional frequency shifts and mode mixing through the matrix elements
\begin{equation}
U_{\nu\nu'}
=
\frac{1}{V}\int_V d^3r\,
\mathbf m_{\nu'}^\dagger(\mathbf r)\cdot\mathcal{H}_{\rm{IFE}}\cdot
\mathbf m_{\nu}(\mathbf r).
\label{eq:gIFE_definition_collinear}
\end{equation}
The diagonal elements $U_{\nu\nu}$ describe the IFE-induced line shifts, whereas the off-diagonal elements $U_{\nu\nu'}$ with $\nu\neq \nu'$ describe optically induced coupling. In particular, the matrix element $U_{000}^{100}$ is nonzero because the IFE field $H_z^{\rm IFE}$ is both parity-breaking and spatially inhomogeneous.

Passing to the frequency domain, $c_\nu(t)\propto e^{-i\Omega t}$, Eq.~\eqref{eq:CMT_collinear_time} reduces to the two-mode eigenvalue problem
\begin{equation}
\begin{pmatrix}
\widetilde\Omega_{000} + U_{000} & U_{000}^{100}\\
\left(U_{000}^{100}\right)^\ast & \widetilde\Omega_{100} + U_{100}
\end{pmatrix}
\begin{pmatrix}
c_{000}\\
c_{100}
\end{pmatrix}
=
\Omega
\begin{pmatrix}
c_{000}\\
c_{100}
\end{pmatrix},
\label{eq:CMT_2x2_collinear}
\end{equation}
whose eigenfrequencies are
\begin{equation}
\Omega_\pm=
\overline{\Omega}
\pm
\sqrt{
\delta\widetilde\Omega^2
+
|U_{000}^{100}|^2
},
\label{eq:Omega_pm_collinear}
\end{equation}
with
\begin{equation*}
\begin{split}
\overline{\Omega} &=
\frac{\widetilde\Omega_{000}+U_{000}+\widetilde\Omega_{100}+U_{100}}{2},\\
\delta\widetilde\Omega &=
\frac{\widetilde\Omega_{000}+U_{000}-\widetilde\Omega_{100}-U_{100}}{2}.
\end{split}
\end{equation*}

This form explicitly shows the avoided crossing induced by the optical pump. The eigenlines obtained from the CMT model are plotted as solid lines in Fig.~\ref{fig:Figure_3}(a). The agreement with the numerical COMSOL results is excellent. The dotted lines in Fig.~\ref{fig:Figure_3}(a) show the diagonal-only approximation, i.e. shifted but non-hybridizing eigenlines, which recover the asymptotic behavior far from the anticrossing.

The CMT formulation also makes it possible to extract the dependence of the spectrum on pump and material parameters such as intensity, wavelength, and gyration factor. To this end, we rewrite Eq.~\eqref{eq:H_IFE} for the effective IFE field as
\[
H_z^{\rm IFE}
=
-\varepsilon'(\lambda)\frac{\varepsilon_0}{\mu_0}\frac{g(\lambda)}{M_s}
|\mathbf E_0|^2
\mathcal S(\mathbf r),
\]
where $\mathcal{S}(\mathbf r)$ is the dimensionless $z$-component of the optical spin density, resulting from bilinear combinations of the terms in Eq.~\eqref{eq:Mie_internal}. The matrix elements can then be written as
\begin{equation*}
U_{\nu\nu'}
=
2\gamma\frac{\varepsilon'(\lambda)g(\lambda)}{\mu_0 c M_s}\,
I\,u_{\nu\nu'},
\end{equation*}
with
\[
u_{\nu\nu'}
=
\frac{1}{V}\int_V d^3r\,
\mathcal{S}(\mathbf r)\,
\mathbf{m}_{\nu'}^\dagger(\mathbf r)\cdot\hat{S}_z\cdot
\mathbf{m}_{\nu}(\mathbf r).
\]
This representation makes the dependence on pump parameters explicit. In particular, at exact resonance of the unperturbed modes, $\widetilde\Omega_{000}=\widetilde\Omega_{100}$, the splitting is given by
\begin{equation}
\Delta\Omega
=
\frac{4\gamma\varepsilon'g}{\mu_0 c M_s}\,
I\,
\sqrt{
\left(
\frac{u_{000}-u_{100}}{2}
\right)^2
+
|u_{000}^{100}|^2
}.
\label{eq:gap_collinear}
\end{equation}
This expression shows explicitly that the splitting is linear in the pump intensity, in agreement with the numerical dependence shown in Fig.~\ref{fig:Figure_3}(c).
Equation~\eqref{eq:gap_collinear} also clarifies why the maximum of $\Delta\Omega(\lambda)$ need not coincide exactly with the maximum of the scattering cross section or with the maximum of the volume-averaged field-enhancement factor $f$. 
The splitting is proportional not to $\langle |\mathbf E_{\rm in}|^2\rangle$ itself, but to the symmetry-selected overlap of the magnon modes with
\[
    \varepsilon'(\lambda)g(\lambda)
    i\left(\mathbf E_{\rm in}^{\ast}\times\mathbf E_{\rm in}\right)_z .
\]
Thus, the position of the maximum is affected by the wavelength dependence of the Faraday gyration, the real part of the permittivity, and, most importantly, by the resonant spatial distribution and sign structure of the $z$-component of the optical spin density inside the sphere.
The remaining material parameters mainly affect the magnetic part of the problem.
The saturation magnetization $M_s$ enters both the dipole--exchange spectrum and the IFE prefactor.
The exchange stiffness $A$ enters through $l_{\rm exch}=\sqrt{A/M_s}$ and therefore shifts the radius or field at which the unperturbed modes cross.
The Gilbert damping $\alpha$ determines the linewidths and hence the resolvability of the splitting, but does not shift the lossless avoided crossing to leading order.
Hence, the splitting is linear in the pump intensity and reaches its largest values near the optical resonance, where both the gyration factor $g(\lambda)$ and the mode-overlap coefficients are favorable. 
Equation~\eqref{eq:gap_collinear} can also be used to evaluate the wavelength dependence of the splitting. In that case, however, the dependence is distributed among the dispersive factors $\varepsilon'(\lambda)$ and $g(\lambda)$, as well as the dimensionless overlap coefficients $u_{\nu\nu'}$, which encode the resonant optical-field structure. The corresponding CMT result is shown as the solid line in Fig.~\ref{fig:Figure_3}(b).

As follows from the same coupled-mode expressions, not only the splitting but also the overall frequency shifts are linear in the pump intensity, since the diagonal IFE matrix elements $U_{\nu\nu}$ are proportional to $H_z^{\rm IFE}\propto I$. For a representative intensity $I \approx 5.3\times10^5\,\mathrm{W/cm^2}$,
the present system exhibits an overall frequency shift of the order of $\sim 10\,\mathrm{MHz}$. The direction of this shift depends on the sign of the effective inverse-Faraday field $H_z^{\rm IFE}$, and therefore on the illumination direction, since reversing the propagation direction changes the sign of the induced optical spin and hence of the corresponding Zeeman-like perturbation.

Thus, in the collinear geometry, interference between the ED and MD Mie fields generates an axisymmetric but parity-breaking IFE perturbation, which hybridizes the Kittel mode with the first odd magnon mode and produces an avoided crossing whose magnitude is controlled directly by the pump intensity.

\section{Discussion and outlook}
\label{sec:discussion}

The results above illustrate only the simplest realization of optically induced magnon hybridization, namely the coupling of two modes enabled by parity breaking in the collinear-incidence geometry. In this case, the inverse Faraday effect preserves axial symmetry while breaking mirror symmetry with respect to the equatorial plane, which is sufficient to activate coupling between the Kittel mode and the first odd mode. More generally, however, the hybridization pattern is dictated by the symmetries of the combined magnonic and optical system. The allowed and forbidden couplings can therefore be identified systematically by analyzing which conserved quantities of the unperturbed resonator remain intact under optical driving and which are broken by the perturbation.

This viewpoint opens a broader route toward engineering magnon spectra with light.
The symmetry-protected crossing is exact for the ideal uniformly magnetized sphere considered here.
Static imperfections, such as shape asymmetry, surface roughness, material inhomogeneity, or nonuniform equilibrium magnetization, may break the relevant symmetry and produce residual splitting.
Such sample-specific effects are distinct from the optically tunable IFE coupling and can be included in the same coupled-mode framework as additional perturbations.
By designing the optical excitation to break selected symmetries, one can in principle target different classes of magnon-mode couplings and thereby establish the corresponding selection rules. 
In particular, while the collinear geometry considered here preserves axial symmetry, oblique incidence breaks axial symmetry more strongly by exciting additional azimuthal channels, so couplings between different $m$ or $\widehat{J}_z$ sectors can become allowed.
The same symmetry-based CMT framework is material-independent and can be applied, for example, to YIG, although different values of $M_s$, $A$, damping, optical loss, and gyration shift the spin-wave crossings and require different pump wavelengths or intensities. 
Changing the helicity from $\sigma=+1$ to $\sigma=-1$ reverses the optical spin density and hence changes the sign of the IFE-induced Zeeman-like perturbation, while linearly polarized excitation does not provide the same collinear IFE channel and would require analyzing the local optical spin generated by the specific internal Mie-field structure.
More generally, varying the illumination geometry, polarization state, or resonant optical-mode composition provides a flexible toolbox for symmetry-controlled hybridization of spin-wave modes, extending well beyond the specific parity-breaking mechanism analyzed in this work.

From the experimental standpoint, the key question is not only the magnitude of the optically induced splitting, but also whether it exceeds the spin-wave linewidth and is therefore spectroscopically resolvable. Within the coupled-mode description, the linewidth can be incorporated naturally by retaining the Gilbert damping term in the linearized Landau--Lifshitz--Gilbert equation. In the weak-damping limit, $\alpha\ll 1$, which is consistent with the low microwave loss reported for BIG \cite{10.1063/1.1287227}, the dynamics can be written in the non-Hermitian form
\begin{equation*}
-i\partial_t\mathbf m=\mathcal H_{\rm NH}\mathbf m,
\qquad
\mathcal H_{\rm NH}
=
\mathcal H_{\rm eff}
-
i\alpha\,\hat S_z\,\mathcal H_{\rm eff},
\label{eq:H_nonherm}
\end{equation*}

\begin{figure}
  \centering
    \includegraphics[width=1\columnwidth]{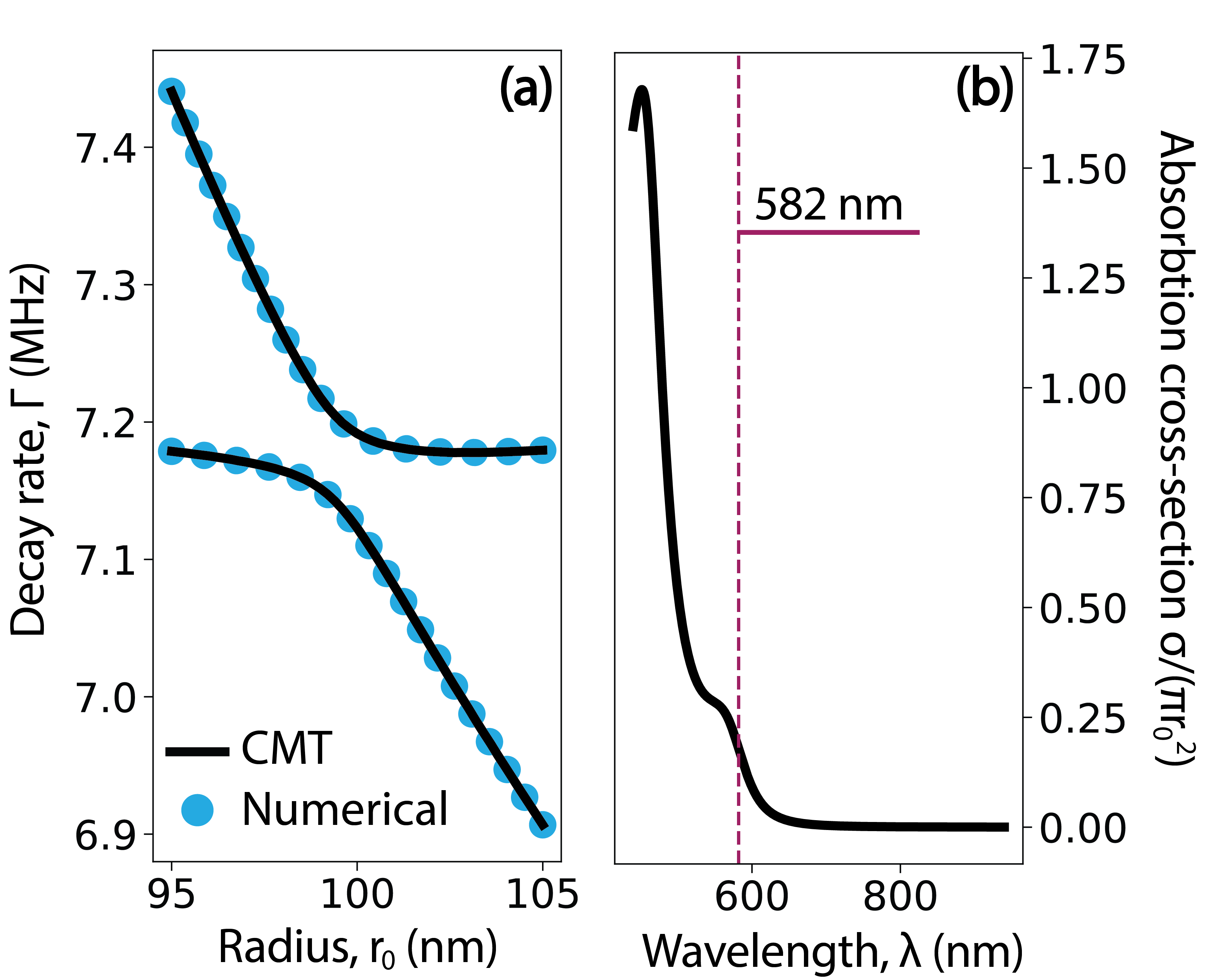}
    \caption{BIG sphere characteristics: (a) decay rates of the hybridized branches for the lossy spin-wave system calculated at pump intensity $I \approx 5.3\times10^5\,\mathrm{W/cm^2}$ and wavelength $\lambda=582 \ \mathrm{nm}$; hollow circles show numerical data, solid lines show the CMT calculations; (b) normalized absorption cross-section, calculated for the same intensity at the sphere radius $r_0=99 \ \mathrm{nm}$; dashed line marks pump wavelength.}\label{fig:Figure_4}
\end{figure}

where $\mathcal H_{\rm eff}=\mathcal H_{\rm exch}+\mathcal H_{\rm dip}+\mathcal H_{\rm IFE}$. Projecting this equation onto the same exchange-mode basis as in the Hermitian CMT problem yields a complex coupled-mode matrix,

\begin{equation*}
H^{\rm CMT}_{\mu\nu}
\;\to\;
H^{\rm CMT}_{\mu\nu}
-
i\alpha D_{\mu\nu},
\label{eq:CMT_nonherm_compact}
\end{equation*}
with
\begin{equation*}
D_{\mu\nu}
=
\int_V d^3r\,
\mathbf m_\mu^\ast(\mathbf r)\,
\hat S_z\mathcal H_{\rm eff}\,
\mathbf m_\nu(\mathbf r).
\label{eq:Dmn}
\end{equation*}
Thus, damping enters as an anti-Hermitian correction to the same matrix that governs coherent hybridization. For an isolated mode, this simply shifts the eigenfrequency into the complex plane and defines its linewidth through the corresponding diagonal damping matrix element. In the two-mode anticrossing problem, one instead diagonalizes the full non-Hermitian CMT matrix and obtains complex eigenfrequencies $\Omega_{\nu}^{\rm NH}=\Omega_{\nu}-i\Gamma_{\nu}$, where the imaginary parts $\Gamma_{\nu}$ determine the decay rates of the hybridized branches. In this notation, $\Gamma_{\nu}$ is the modal decay rate, whereas the corresponding full linewidth is $2\Gamma_{\nu}$ in angular-frequency units, or $2\Gamma_{\nu}/2\pi$ when expressed in frequency units. For the Kittel and $(1,0,0)$ modes, these analytically calculated decay rates are compared with the corresponding numerical results in Fig.~\ref{fig:Figure_4}(a).
This provides a direct criterion for observability: the splitting is experimentally resolvable when the optically induced gap exceeds the relevant linewidths of the participating modes.
Said resolvability criterion is weaker than the strict strong-coupling condition, which requires the coherent coupling rate to exceed the relevant loss rates; here Fig.~3(c) is therefore used primarily as a benchmark for observing a resolved avoided crossing in frequency-domain spectra.

In this respect, Fig.~\ref{fig:Figure_3}(b) supports the experimental plausibility of the effect. For pump intensities in the range $10^4$--$10^6\,\mathrm{W/cm^2}$, the predicted splittings already reach the MHz to hundreds-of-MHz range, while GHz splittings may become accessible at somewhat higher pump intensities under favorable focusing and thermal-management conditions. As illustrated by Fig.~\ref{fig:Figure_3}(c), these values become comparable to, and can exceed, the estimated linewidth of the participating spin-wave modes, indicating that the hybridization should become spectroscopically resolvable once the pump intensity exceeds the corresponding threshold, provided that damping and heating remain sufficiently controlled.

At the same time, thermal management is essential, since even weak absorption can produce a finite steady-state temperature rise under continuous-wave pumping.
Writing $P_{\rm abs}=I C_{\rm abs}=I Q_{\rm abs}\pi r_0^2$, where $Q_{\rm abs}=C_{\rm abs}/\pi r_0^2$, gives $P_{\rm abs}\simeq 1.6\times 10^{-4}Q_{\rm abs}\,{\rm W}$ for $I=5.3\times 10^5\,{\rm W/cm^2}$ and $r_0\simeq 99\,{\rm nm}$.
Approximating the particle as a steady-state nanoscale heat source in a medium with thermal conductivity $\kappa_{\rm th}$ yields
\[
    \Delta T \approx \frac{P_{\rm abs}}{4\pi\kappa_{\rm th}r_0}
    =\frac{I Q_{\rm abs}r_0}{4\kappa_{\rm th}}
\]
For a representative $Q_{\rm abs}=0.1$, an air-suspended particle with $\kappa_{\rm th}\simeq 0.026\,{\rm W\,m^{-1}K^{-1}}$ would experience a large temperature rise and is therefore not favorable for continuous-wave operation~\cite{Huber2011CRC}.
In contrast, fused silica with $\kappa_{\rm th}\simeq 1.38\,{\rm W\,m^{-1}K^{-1}}$ gives $\Delta T\sim 9\,{\rm K}$~\cite{CrystranFusedSilica}, while sapphire with $\kappa_{\rm th}=27.21\,{\rm W\,m^{-1}K^{-1}}$ gives $\Delta T\lesssim 0.5\,{\rm K}$~\cite{CrystranSapphire}.

The substrate also changes the optical environment.
BIG has a high refractive index, with reported values $n_{\rm BIG}=2.819$ at $633\,{\rm nm}$ and $n_{\rm BIG}=2.584$ at $1550\,{\rm nm}$~\cite{Tepper2003BIG}, so silica or glass with $n\simeq 1.46$ preserves strong dielectric contrast~\cite{Malitson1965SiO2,CrystranFusedSilica}, while sapphire provides stronger heat sinking and still retains moderate optical contrast, with refractive index around $n\simeq 1.75$ in the near-infrared/visible range~\cite{CrystranSapphire}.
Here we use the ideal isolated sphere as a proof-of-concept setting that exposes the symmetry mechanism and benchmarks the CMT; for a specific substrate, the same formalism can be applied after recalculating the modified Mie field and IFE overlap integrals.

Heating may shift the spectrum through the temperature dependence of $M_s$ and anisotropy, and may increase linewidths through Gilbert damping.
For high-thermal -conductivity substrates, the estimated $\Delta T$ is sub-Kelvin to few-Kelvin, so thermal shifts should remain smaller than the predicted MHz--hundreds-of-MHz optically induced splittings in Fig.~\ref{fig:Figure_3}.
Poorly heat-sunk particles, however, would require pulsed excitation, reduced duty cycle, or active cooling.

A further route toward experimental implementation is to use ensembles of nominally identical particles rather than a single resonator. In such an arrangement, each particle can be treated as an individual spin-wave resonator, while the interparticle interaction is mediated primarily by the dipolar stray field. Within the same coupled-mode picture, the interaction between a mode $\mu$ of one particle and a mode $\nu$ of another is described by an off-diagonal matrix element of the form
\begin{equation*}
K_{\mu\nu}^{(12)}
\sim
\gamma M_s
\int_{V_1} d^3r
\int_{V_2} d^3r'\,
\mathbf m_\mu^\dagger(\mathbf r)\cdot
\hat S_z \hat K(\mathbf r-\mathbf r')\,
\mathbf m_\nu(\mathbf r'),
\end{equation*}
where $\hat K(\mathbf r-\mathbf r')$ is the magnetostatic dipolar kernel. For two particles separated by a center-to-center distance $d$ much larger than their radius, one may approximate $|\mathbf r-\mathbf r'|\approx d$ across the integration domain, thus $K_{\mu\nu}^{(12)}\propto d^{-3}$ as the dipolar kernel decays as $|\mathbf{r}-\mathbf{r}'|^{-3}$, as can be derived from Eq.~\eqref{eq:K_def}.
Thus, for sufficiently dilute arrays, the dipolar coupling between particles decreases rapidly with separation and can be made much smaller than the intrinsic mode frequencies, linewidths, and optically induced splittings. In that regime, the particles act to a good approximation as independent spin-wave resonators, so the measured response is simply the sum of many nearly uncoupled contributions, which may be advantageous for signal enhancement.

More generally, the symmetry-based mechanism discussed here is not restricted to spheres. Other resonator geometries, such as finite cylinders, are also natural candidates, since they likewise possess axial symmetry and a well-defined mirror parity in the unperturbed problem. As shown in Ref.~\cite{shuklin2026dipoleexchangespinwavesmode}, these symmetries remain the natural organizing principles for the dipole--exchange spectrum in cylindrical particles as well, making them a promising platform for extending the present analysis to optically induced mode hybridization beyond spherical geometry.

\section{Conclusions}

We have shown that internal Mie-resonant inverse-Faraday fields can be used to engineer the dipole--exchange spin-wave spectrum of a BIG sphere through symmetry-selective optical perturbations. In the collinear-incidence geometry, the optical drive remains axisymmetric but breaks mirror parity, which enables coupling between the Kittel mode and the first odd dipole--exchange mode while preserving the $\widehat{J}_z$ sector structure. Within coupled-mode theory, this hybridization is described by a simple two-mode model that reproduces the numerically obtained avoided crossing and yields analytical expressions for the induced splitting and its dependence on pump parameters.

The resulting splitting is linear in pump intensity and is enhanced near optical Mie resonances, where both the internal field enhancement and the magneto-optical response are favorable. Estimated linewidths, together with the calculated MHz--hundreds-of-MHz splittings, indicate that the effect should be observable under realistic experimental conditions, provided that thermal loading is properly managed. More broadly, the present results establish symmetry analysis as a practical route toward optical control of magnon spectra in confined ferrimagnetic resonators and suggest natural extensions to other incidence geometries and resonator shapes.

\section{Funding}
This work was supported by Russian Scientific
Foundation (project No. 24-72-00028).











\printcredits

\bibliographystyle{unsrt}
\bibliography{references}



\end{document}